%% file: ms.tex
\documentclass[12pt]{iopart}
\pdfoutput=1 
\usepackage[T1]{fontenc}
\usepackage[colorlinks=true,unicode=true,allcolors=blue]{hyperref}
\usepackage{graphicx} 
\usepackage{booktabs}
\usepackage{siunitx}
\usepackage[USenglish]{babel}
\usepackage{csquotes}

\begin{document}

\title[Atom Cloud Detection Using a Deep Neural Network]{Atom Cloud Detection Using a Deep Neural Network}

\author{Lucas R. Hofer\textsuperscript{\normalfont 1}, Milan Krstaji\'{c}\textsuperscript{\normalfont 1,2}, 
P\'{e}ter Juh\'{a}sz\textsuperscript{\normalfont 1}, Anna L. Marchant\textsuperscript{\normalfont 1} and Robert P. Smith\textsuperscript{\normalfont 1}}

\address{\textsuperscript{1} Clarendon Laboratory, University of Oxford, Parks Road, Oxford OX1 3PU, United Kingdom}
\address{\textsuperscript{2} Cavendish Laboratory, University of Cambridge, J. J. Thomson Avenue, Cambridge CB3 0HE, United Kingdom}
\ead{robert.smith@physics.ox.ac.uk}
\input{abstract}

%
\vspace{2pc}

\input{body}

\section*{References}
\bibliographystyle{iopart-num}
\input{output.bbl}



\end{document}

%% file: abstract.tex
\begin{abstract}
We use a deep neural network to detect and place region-of-interest boxes around ultracold atom clouds in absorption and fluorescence images---with the ability to identify and bound multiple clouds within a single image. The neural network also outputs segmentation masks that identify the size, shape and orientation of each cloud from which we extract the clouds' Gaussian parameters. This allows 2D Gaussian fits to be reliably seeded thereby enabling fully automatic image processing.  
\end{abstract}


%% file: body.tex
\section{Introduction}

\begin{figure}[h]
\centering\includegraphics[width=.8\linewidth]{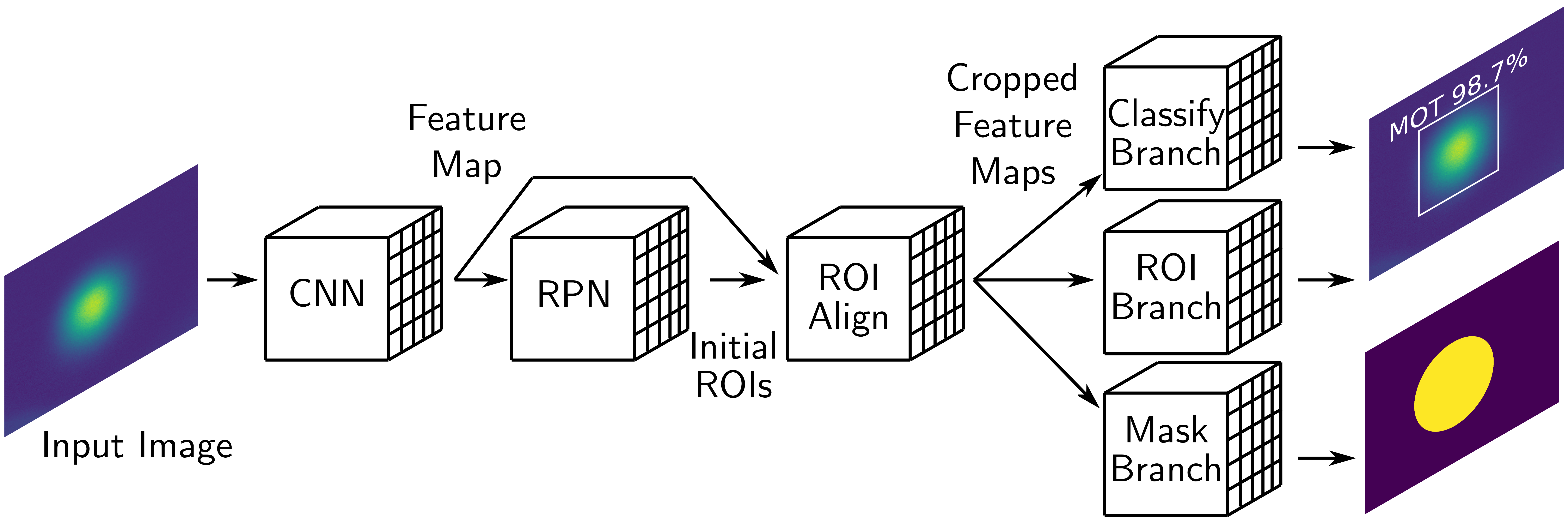}
\caption{Overview of the neural network (NN). An experimental image is initially fed into a convolutional neural network (CNN) which produces a `feature map' of the image. The region-proposal network (RPN) uses this feature map to generate regions-of-interest (ROI) where atom clouds are likely contained. In the ROI alignment stage, the CNN's feature map is cropped and resized for each cloud's ROI before being fed into three parallel branches. The first branch (ROI) generates refined ROIs, whereas the second branch (class) classifies the type of atom cloud in the ROI. Finally, the third branch (mask) generates a segmentation mask corresponding to the $1/e^2$ contour of the atom cloud.}
\label{fig:overview}
\end{figure}

Deep neural networks have revolutionized data analysis and led to automation of tasks that previously required human supervision. Image analysis has particularly benefited through the use of convolutional neural networks (CNNs) \cite{lecun1998gradient} and their derivatives which have allowed for image classification \cite{krizhevsky2012imagenet, simonyan2014very}, object detection \cite{girshick2014rich, redmon2016you} and instance segmentation \cite{long2015fully}. Although many of these neural networks (NNs) were developed for tasks such as facial recognition by social media networks \cite{taigman2014deepface, schroff2015facenet}, they have also been used to  identify laser modes \cite{Hofer2019}, classify phases in condensed-matter systems \cite{ch2017machine, carrasquilla2017machine}, reduce measurement errors for trapped-ion qubits \cite{seif2018machine} and process images from cold atom experiments \cite{picard2019deep, ness2020single, miles2020correlator}. In this work, we use an instance segmentation NN (see Fig.~\ref{fig:overview}) to analyze experimental images containing atom clouds in magneto-optical traps (MOTs) and optical dipole traps (ODTs).

Neural networks consist of an input layer and an output layer with a number of intermediate hidden layers which are connected to one another via `weights'. Rather than employing hard-coded algorithms, NNs learn to emulate data they encounter through training cycles, in which data is iteratively passed through the NN and the output compared to the `ground truth'. The difference between the two is then used to update the weights between the NN's layers, thereby improving its accuracy. When employing supervised training, this requires a dataset which includes both input data and their associated ground truth values. For object detection NNs, these ground truth values are rectangular bounding boxes for each object, as well as labels classifying the object types in the bounding boxes. Instance segmentation NNs build upon object detection NNs by also requiring pixel-to-pixel segmentation masks in which image pixels comprising the object have mask values of one, whereas all other pixels have mask values of zero. 

\begin{figure}[t!]
\centering\includegraphics[width=\linewidth]{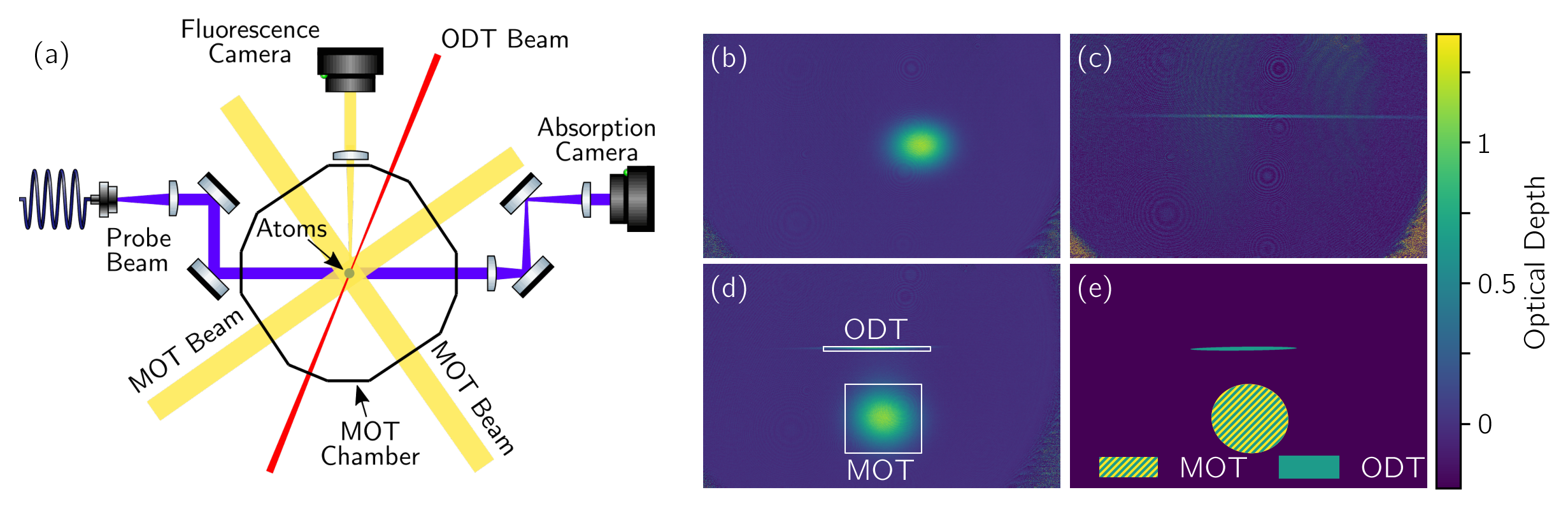}
\caption{Construction of labelled experimental dataset. (a) Experimental setup. The MOT is formed from six \SI{583}{\nano\metre} laser beams in conjunction with a quadrupole magnetic field. Atoms trapped in the MOT can be loaded into an ODT formed from a single \SI{1030}{\nano\metre} laser beam. The fluorescence camera captures the light scattered by the atoms from the MOT beams. During absorption imaging, a \SI{401}{\nano\metre} probe beam passes through the atom cloud and absorbs a portion of the light creating a shadow which is imaged by the camera. (b) Image of an atom cloud released from the MOT. (c) Image of an atom cloud released from the ODT. The optical depth is multiplied by six to increase ODT visibility. (d) Image with both MOT and ODT atom clouds. Each cloud is labelled and a region-of-interest (ROI) box is drawn around it (white lines). (e) Segmentation masks for the MOT (hatched yellow) and ODT (solid green) clouds in (d).}
\label{fig:dataset}
\end{figure}

Our dataset consists of  images of cold atom clouds in a MOT \cite{raab1987trapping} and an ODT \cite{chu1986experimental} (see Fig.~\ref{fig:dataset}a-c). Atom clouds in these traps form approximately Gaussian density distributions \cite{becbook}. Fitting a cloud allows the parameters describing the distribution (Gaussian parameters) to be extracted and used to ascertain information such as the cloud's size and density. Furthermore, by using time-of-flight measurements \cite{lett1988observation} the temperature of the atoms can be determined.

A region-of-interest (ROI) \cite{muessel2013optimized} centered on the atom cloud is used during fitting as objects in the image other than the atom cloud can cause an inaccurate fit (e.g. an atom cloud in another trap or extraneous noise). Additionally, decreasing the fit area can significantly decrease the fit time when using two-dimensional fitting. Manually determining the ROI is time-consuming when analyzing a large number of images and an algorithmic procedure is often employed \cite{barker2020applying}, such as finding the highest intensity region of the image or taking the `center of mass' and then iteratively expanding the ROI around this point until the fractional enclosed `power' exceeds some threshold. However, if the image is noisy (e.g. contains fringing), the ROI will be inaccurate. Additionally, these types of algorithms are not applicable to images with multiple atom clouds (see Fig.~\ref{fig:dataset}d).

Here we propose a deep neural network based approach to ROI determination in which a NN finds the ROI for each atom cloud in an image (see Fig.~\ref{fig:overview}). Furthermore, the NN differentiates between clouds in a MOT and those in an ODT and also outputs a segmentation mask for each cloud from which Gaussian parameters are directly extracted. 

The rest of the article is arranged as follows: Sec.~\ref{sec:dataset} describes the experimental dataset used for NN training and validation, Sec.~\ref{sec:nn} describes the training process and Sec.~\ref{sec:bo} discusses Bayesian optimization \cite{shahriari2015taking, snoek2012practical} of the NN's hyperparameters \cite{snoek2012practical}. Lastly,  Sec.~\ref{sec:gaussian} examines how the Gaussian parameters are calculated from the segmentation mask.

\section{Experiment and Dataset}\label{sec:dataset}

To produce the ultracold atom clouds, erbium atoms are initially trapped and cooled to $\sim$\SI{20}{\micro\kelvin} in a narrow-line MOT \cite{frisch2012narrow}  before being loaded into an ODT formed from a \SI{30}{\watt}, \SI{1030}{\nano\metre} laser beam focused down to a $\sim$\SI{40}{\micro\metre} waist (see Fig.~\ref{fig:dataset}).
Optimization of the trap loading involves maximizing the  atom number while minimizing the cloud temperature. The atom number is found by fitting the atom cloud in either a fluorescence or absorption image\textsuperscript{\footnotemark}\footnotetext{Experimental images shown in the paper are processed from 2--3 raw images. Fluorescence images require both an image with atoms and a background image without atoms. Absorption images additionally require a probe image in which the probe beam is turned on, but no atoms are present.} with a two-dimensional (2D) Gaussian (see Eq.~\ref{eq:2dgauss}) and then integrating under the curve.  The cloud temperature can be determined from how the cloud width evolves during time-of-flight expansion.

The experimental dataset consists of 260 fluorescence and absorption images along with their ROIs, labels and segmentation masks. Of these, 130 images contain clouds released from the MOT with no ODT present (see Fig.~\ref{fig:dataset}b) and 130 images contain either just atoms released from the ODT (see Fig.~\ref{fig:dataset}c) or images where atoms released from the ODT and the MOT are both present (see Fig.~\ref{fig:dataset}d). We manually label the atom clouds in the images as `MOT' or `ODT' and draw a ROI box at the clouds' edges which we define as the 1/$e^2$ radii along the $x$ and $y$ axes (see Fig.~\ref{fig:dataset}d). This definition prevents the ROI boxes from overlapping when the MOT and ODT are both present; however, the ROI boxes are also easily expandable when analysis requires the wings of the distribution.

The manually drawn ROIs were expanded by a factor of two---excepting where the expanded ROIs would overlap---and the atom clouds inside fit with a 2D Gaussian
\begin{equation}
I(x, y) = I_{\text{b}}+I_0 e^{-2\left(\frac{\left[\left(x-x_0\right)\cos\theta+\left(y-y_0\right)\sin\theta\right]^2}{w_x^2}+\frac{\left[\left(y-y_0\right)\cos\theta-\left(x-x_0\right)\sin\theta\right]^2}{w_y^2}\right)}\text{,}
\label{eq:2dgauss}
\end{equation}
\noindent where $I(x, y)$ is the image intensity, $I_{\text{b}}$ is the background intensity, $I_0$ is the peak intensity, $x_0$ and $y_0$ are the center coordinates,  $w_x$ and $w_y$ are the 1/$e^2$ radii along the major and minor axes and $\theta$ is the angular orientation of the distribution. To increase the accuracy of the ROIs used for training, the fit parameters were used to calculate the 1/$e^2$ radii along the image axes \cite{Hofer2019} (previously estimated by eye) and the ROI boxes redrawn using these values. The process of fitting and redrawing the ROI boxes from the fit parameters was then completed a second time with subsequent iterations neglected due to an insignificant increase in accuracy.

A segmentation mask was generated for each atom cloud (see Fig.~\ref{fig:dataset}e) with the mask borders placed at the 1/$e^2$ contour of the cloud---calculated from the final fit parameters; pixels within the 1/$e^2$ contour were set to one, whereas pixels outside were set to zero. Finally, the dataset was randomly split into a training dataset with 200 images and a validation dataset with 60 images.

\section{Neural Network and Training}\label{sec:nn}

\begin{figure}[t!]
\centering\includegraphics[width=\linewidth]{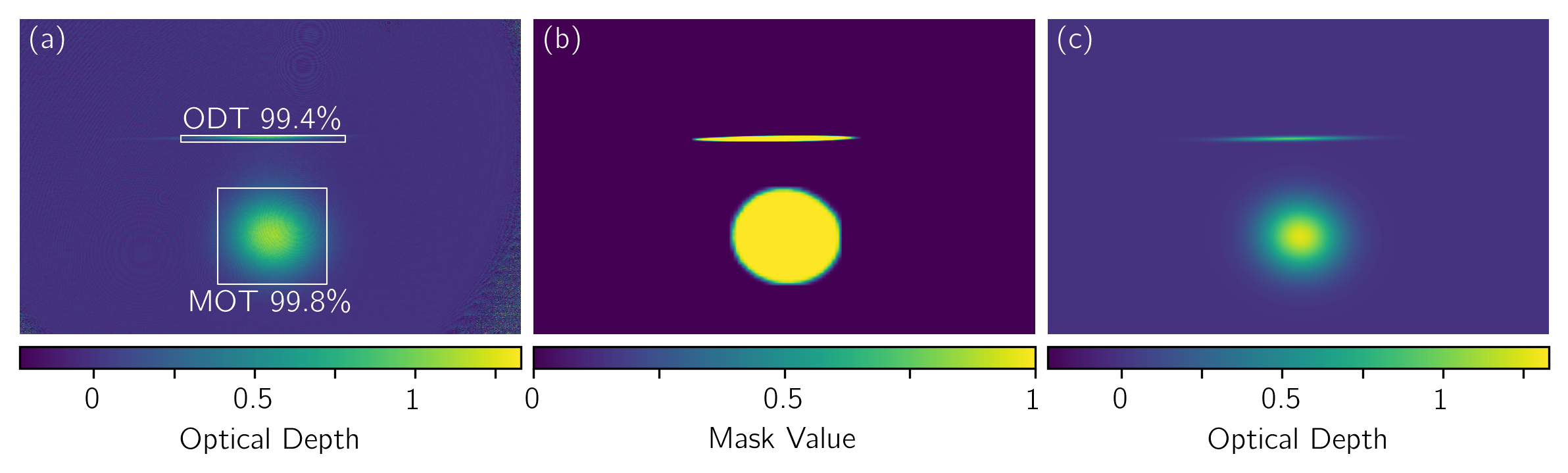}
\caption{Neural network output. (a) Sample image from the validation dataset with the regions-of-interest (solid white lines), labels and confidence scores predicted by the NN. (b) The segmentation masks returned by the NN for (a) which, due to the internal structure of the NN, have values between zero and unity. (c) Reconstruction of (a) using Eq.~\ref{eq:2dgauss} and the extracted 2D Gaussian parameters from (b).}
\label{fig:predict}
\end{figure}

We use the neural network Mask R-CNN \cite{maskrcnn} to detect and bound the atom clouds, as well as to provide segmentation masks for each cloud (see Fig.~\ref{fig:predict}a--b). The NN (see Fig.~\ref{fig:overview}) begins with the experimental image being fed into a convolutional neural network base (CNN, ResNet-50 \cite{resnet}). The CNN's outputted feature map \cite{zeiler2014visualizing} is then input into a region-proposal network (RPN) which generates ROIs where objects are likely located. Next, these ROIs are cropped from the CNN's feature map in a ROI alignment stage and resized to uniform dimensions. The cropped feature maps are then fed into three parallel branches. The first applies a classifier to determine the object type and give the confidence of its prediction---which is helpful in determining whether to use the ROI in post-NN analysis. The second branch gives a more accurate ROI box prediction (see Fig.~\ref{fig:predict}a) and finally the third outputs a segmentation mask for the object inside the ROI (see Fig.~\ref{fig:predict}b). Since all three branches share the same base, computation speed is significantly increased \cite{fasterrcnn}.

\begin{figure}[t!]
\centering\includegraphics[width=\linewidth]{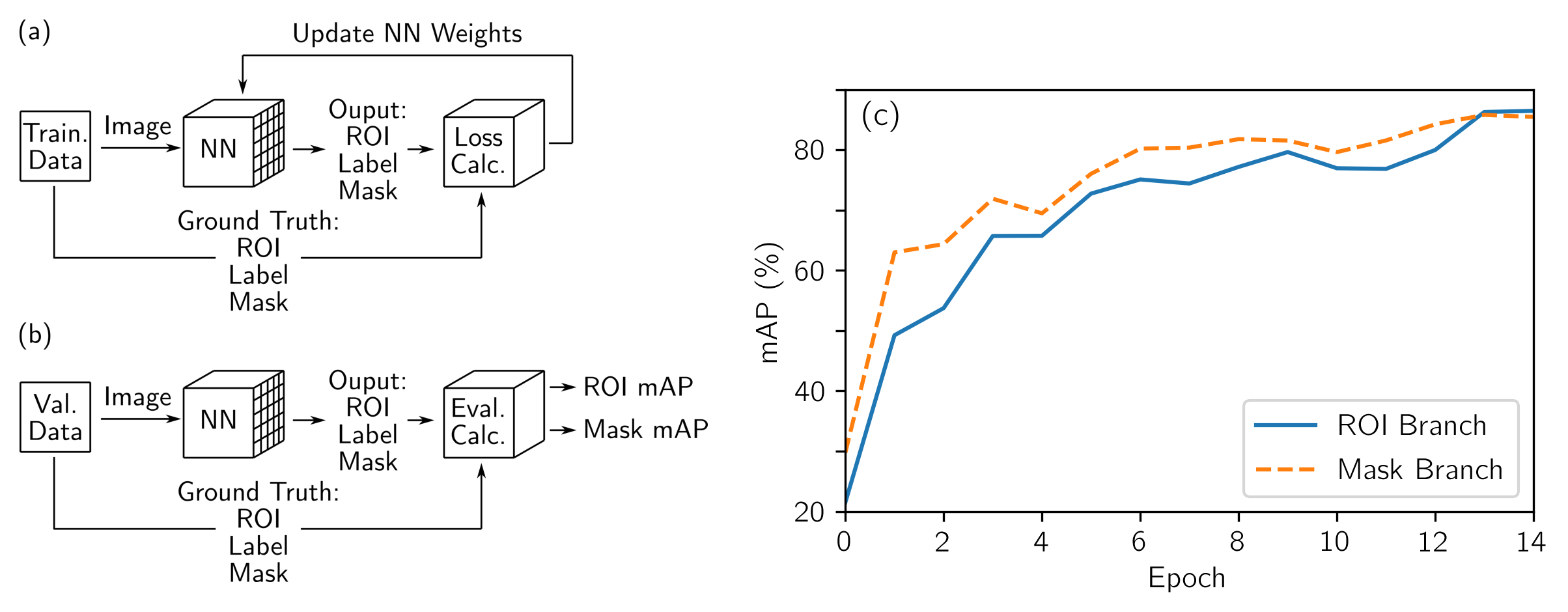}
\caption{Training and validating the neural network. (a) During training of the NN, images from the training dataset are fed into the NN which outputs the regions-of-interest, labels and masks. These outputs are compared to the ground truth values and the loss calculated. The loss is then back-propagated through the NN and used to update the weights between the NN's layers. (b) When evaluating the NN, the validation dataset is used in a process similar to (a); however, the mean average precision (mAP) for both the object detection branch (ROI) and segmentation mask branch are calculated rather than the loss. (c) The mAP of the ROI branch (solid line) and the segmentation mask branch (dashed line) as a function of the training epoch for the NN trained with the best set of hyperparameters (see Table \ref{tab:parameters}).}
\label{fig:training}
\end{figure}

During training, the NN output is compared to the ground truth (i.e. the expected output from the training dataset) via a loss function; the loss is then back-propagated \cite{hecht1992theory} through the NN to adjust the weights between layers and refine the NN model (see Fig.~\ref{fig:training}a). The loss function for the RPN stage is L1 loss  \cite{zhao2016loss}, for the ROI branch it is Smooth L1 loss \cite{feng2018wing}, the classifier uses categorical cross entropy loss \cite{goodfellow2016deep} and lastly the mask branch utilizes binary cross entropy loss \cite{keren2018fast}. Although the loss can be separately back-propagated for each branch, a simpler approach is taken here in which the losses are summed together and then back-propagated to update the weights of the NN \cite{maskrcnn}.

An epoch denotes a single cycle of training in which every image in the training dataset is passed through the NN, the loss calculated and the model weights updated. Increasing the number of training epochs can increase the NN's final accuracy; however, too many epochs can result in overfitting and thus a reduction in the NN's ability to generalize to new data. The NN in this work is trained for fifteen epochs which achieves a high accuracy without overfitting (see Fig.~\ref{fig:training}c) when Bayesian optimization (BO) is utilized to determine the hyperparameters (see Sec.~\ref{sec:bo}). 

Five hyperparameters, which tune the learning process, are set before the NN's training phase. The first is the learning rate which determines the size of the step the NN takes during stochastic gradient descent \cite{bottou1998online}. If the learning rate is too low, the NN will take too long to converge to the minimum of the loss function, whereas if the learning rate is too high, the NN will not be able to descend to the minimum, but will oscillate around it or diverge. Since larger learning rates are useful at the beginning of training and smaller learning rates are useful towards the end, a learning rate scheduler \cite{jastrzkebski2017three} is employed which decreases the learning rate by some scalar (decay, the second hyperparameter) after a fixed number of epochs (step size, the third hyperparameter). The fourth hyperparameter is the momentum which prevents the NN from getting stuck in a local minimum during training \cite{bengio2012practical}. The last hyperparameter is the batch size---the number of images simultaneously passed through the NN---and, unlike the other hyperparameters which are tuned with BO, is fixed to four for all NNs.

The accuracy of the NN is evaluated (see Fig.~\ref{fig:training}b) using the COCO evaluation metric which entails calculating the intersection-over-union \cite{everingham2010pascal},
\begin{equation}
\text{IoU}=\frac{\text{area}\left(\text{ROI}_\text{p}\cap \text{ROI}_\text{gt}\right)}{\text{area}\left(\text{ROI}_\text{p}\cup \text{ROI}_\text{gt}\right)},
\end{equation}
where $\text{ROI}_\text{p}$ is the ROI prediction from the NN and $\text{ROI}_\text{gt}$ is the ground truth ROI. Values below a predetermined IoU threshold (e.g. $\text{IoU}=0.5$) are considered a false prediction---mislabelling the object class within the ROI is also a false prediction---, whereas values above are considered a true prediction. A precision recall-curve \cite{boyd2013area} is then constructed and integrated to give the average precision (AP) for the given threshold (e.g. AP$_{50}$ for the $\text{IoU}=0.5$ threshold). Finally, the AP is calculated for ten IoU thresholds (0.50--0.95 with a step size of 0.05) and averaged to give the mean average precision (mAP)---which is the metric generally reported for object detection. The mask AP values and mAP can be similarly calculated \cite{cordts2016cityscapes}.

The NNs are trained in a Google Colab notebook \cite{bisong2019google} utilizing a GPU backend and are written in Python using the PyTorch package \cite{paszke2017automatic}. Rather than training the NN from scratch, the model weights are loaded from a network pre-trained on the COCO train2017 dataset \cite{lin2014microsoft}---which significantly reduces the time required to train the network and increases the model's final accuracy \cite{yosinski2014transferable}. After each training epoch, the updated NN is evaluated on the validation dataset which yields the mAP for both the object detection and the segmentation mask branches (see Fig.~\ref{fig:training}c). These two values are averaged together to give the accuracy of the training epoch (average mAP). Finally, the epoch with the highest average mAP over the fifteen training epochs determines the overall accuracy of the NN model and the NN weights from this epoch are saved.

\section{Bayesian Optimization of Hyperparameters}\label{sec:bo}

The accuracy of the trained NN is sensitive to the hyperparameters used during training. In the past, the hyperparameters were tuned through grid search, random search \cite{bergstra2012random} or by hand; however, in recent years, Bayesian optimization (BO) has been successfully employed to find the best set of hyperparameters \cite{snoek2012practical, eggensperger2013towards}. BO is particularly useful when trying to find the minimum (or maximum) of a function which is noisy and expensive to evaluate---such as NN training---thereby making a grid search of the parameter space impractical \cite{shahriari2015taking}. 

Bayesian optimization takes the function value (cost) at previously evaluated points and uses a Gaussian process to model the cost as a function of the parameter space \cite{frazier2018tutorial}. The  model also determines the confidence of its predictions in a given region of the parameter space: perfect certainty at evaluated points, low uncertainty near evaluated points and high uncertainty far from evaluated points. The BO loop then determines where to evaluate the function next by weighing the benefits of evaluating the function near the model's predicted minimum (or maximum) or evaluating the function in an unexplored region of the parameter space \cite{brochu2010tutorial}.

\begin{table}[b]
\centering
{\small
\input{hyperparameters}
  \caption{The hyperparameter search space used during Bayesian optimization along with the hyperparameter values used to train the NN to the highest average mAP (see Figs.~\ref{fig:training}a and \ref{fig:contour}). \label{tab:parameters}}}
\end{table}

\begin{figure}[t]
\centering\includegraphics[width=\linewidth]{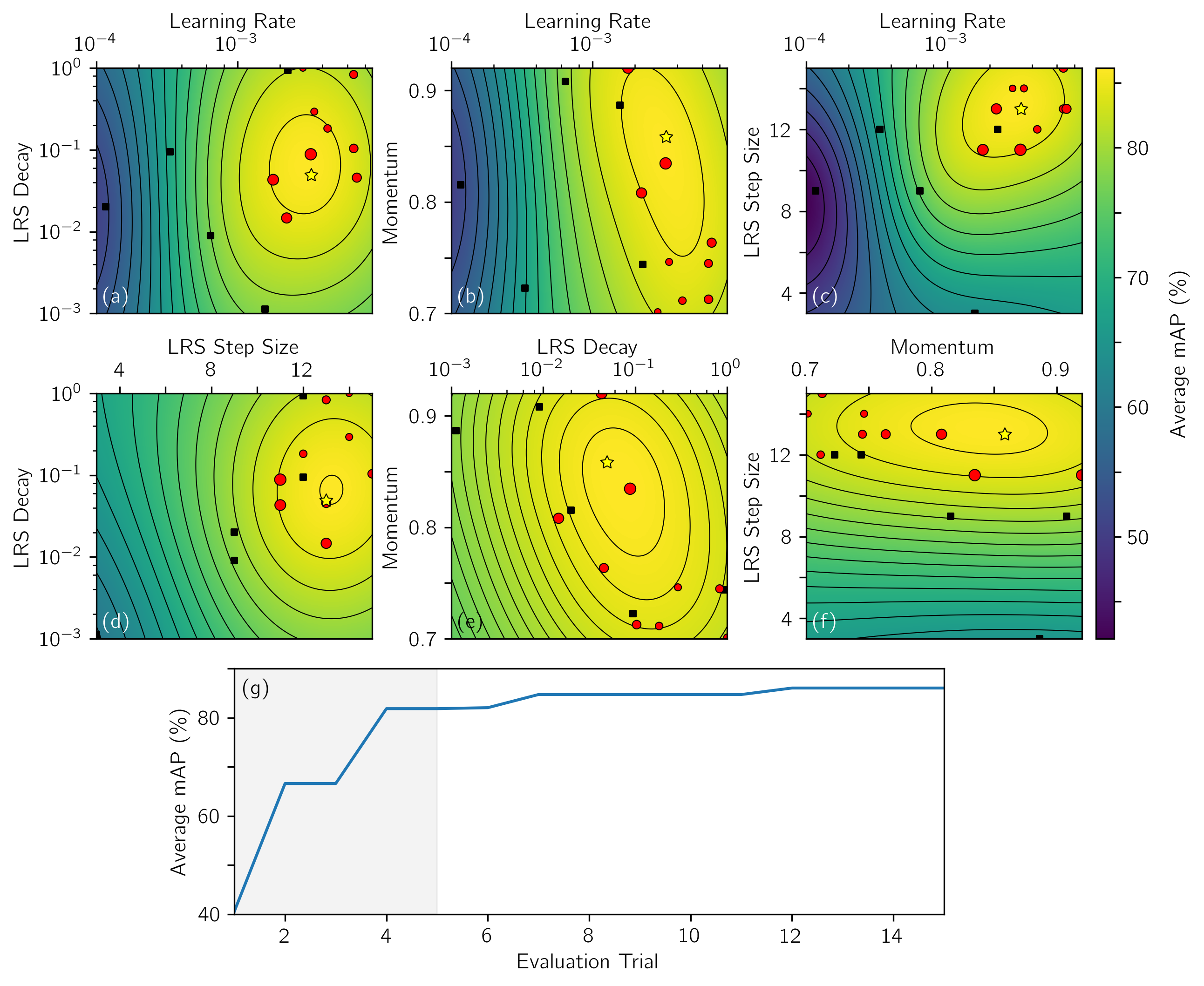}
\vspace{-.85cm}
\caption{Bayesian optimization of hyperparameters. (a)--(f) The average mAP of the NN predicted by the Bayesian optimization (BO) model as a function of the hyperparameters. 
The plots are 2D slices through the 4D parameter space taken at the best position (yellow star) for those parameters not displayed. The NN is initially trained and evaluated at five quasi-random hyperparameter positions (black squares) after which the BO loop iteratively determines the evaluation points (red dots, size increases with trial).  (g) The best achieved average mAP as a function of the evaluation trial; the quasi-random trial region is shaded in grey.}
\label{fig:contour}
\end{figure}

When optimizing our NN training with BO, the average mAP of the NN is evaluated as a function of the hyperparameter space which consists of the learning rate, momentum, and the learning rate scheduler step size and decay (see Table~\ref{tab:parameters}). As a warm start, the NN is initially trained and evaluated at five quasi-randomly  (Sobol generated \cite{sobol1967distribution}) distributed points (see black squares in  Fig.~\ref{fig:contour}a-f). Further evaluation points (red circles in Fig.~\ref{fig:contour}a-f) are iteratively determined by the BO loop. The Ax Python package is used for the BO loop as it provides a high level interface to the BoTorch \cite{balandat2019botorch} BO package.

With an increasing number of BO evaluation trials the best achieved average mAP rises and converges (see Fig.~\ref{fig:contour}g). The best set of hyperparameters (see Table~\ref{tab:parameters}) gives a mAP of 86.3\% for the object detection branch (see Fig.~\ref{fig:training}b) and a mAP of 85.8\% for the mask branch. These mAP values are higher than those for similar NNs trained on the COCO validation dataset \cite{maskrcnn} \cite{redmon2018yolov3} which is likely due to our NN only needing to classify two object types rather than the eighty in the COCO validation dataset \cite{everingham2010pascal}, as well as the relatively simple features of the MOT and ODT clouds.

\section{Gaussian Parameter Analysis}\label{sec:gaussian}

\begin{figure}[t]
\centering\includegraphics[width=.8\linewidth]{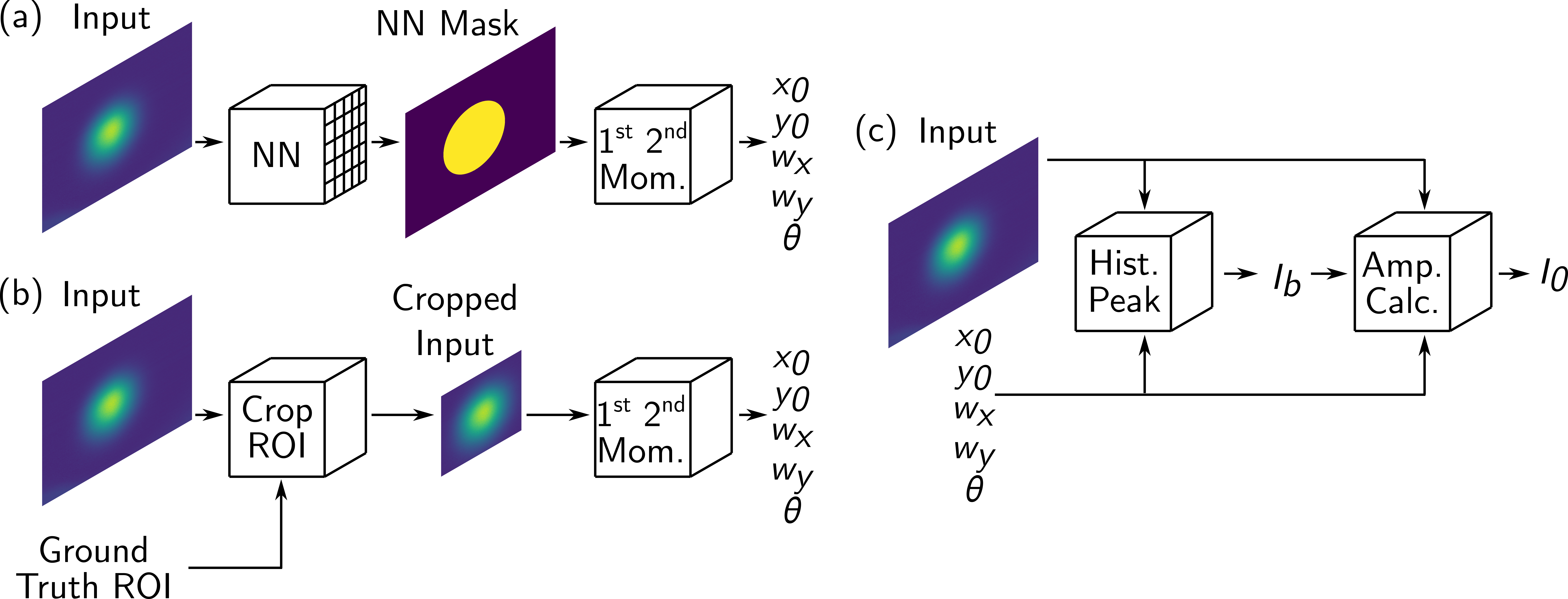}
\caption{Extraction of Gaussian parameters. (a) NN extraction of the parameters is performed by first inputting the experimental image into the NN which returns a segmentation mask for each atom cloud. Applying the first and second moments (mom.) directly to the segmentation mask yields the center coordinates of the atom cloud $\{x_0, y_0\}$, its 1/$e^2$ radii $\{w_x, w_y\}$ and its angular orientation $\theta$.  (b) For conventional extraction the experimental image is cropped at the ROI and the first and second moments applied to the cropped image to yield $\{x_0, y_0, w_x, w_y, \theta \}$. (c) Extracting the intensity parameters. The intensity offset $I_{\text{b}}$ is determined by taking the intensity level corresponding to the peak of a histogram of the experimental image pixel values (hist. peak). The peak intensity $I_0$ is then calculated (see Eq.~\ref{eq:power}) using the power inside the 1/$e^2$ contour and the previously extracted parameters $\{x_0, y_0, w_x, w_y, \theta\}$.}
\label{fig:flow}
\end{figure}

\begin{figure}[t]
\centering\includegraphics[width=\linewidth]{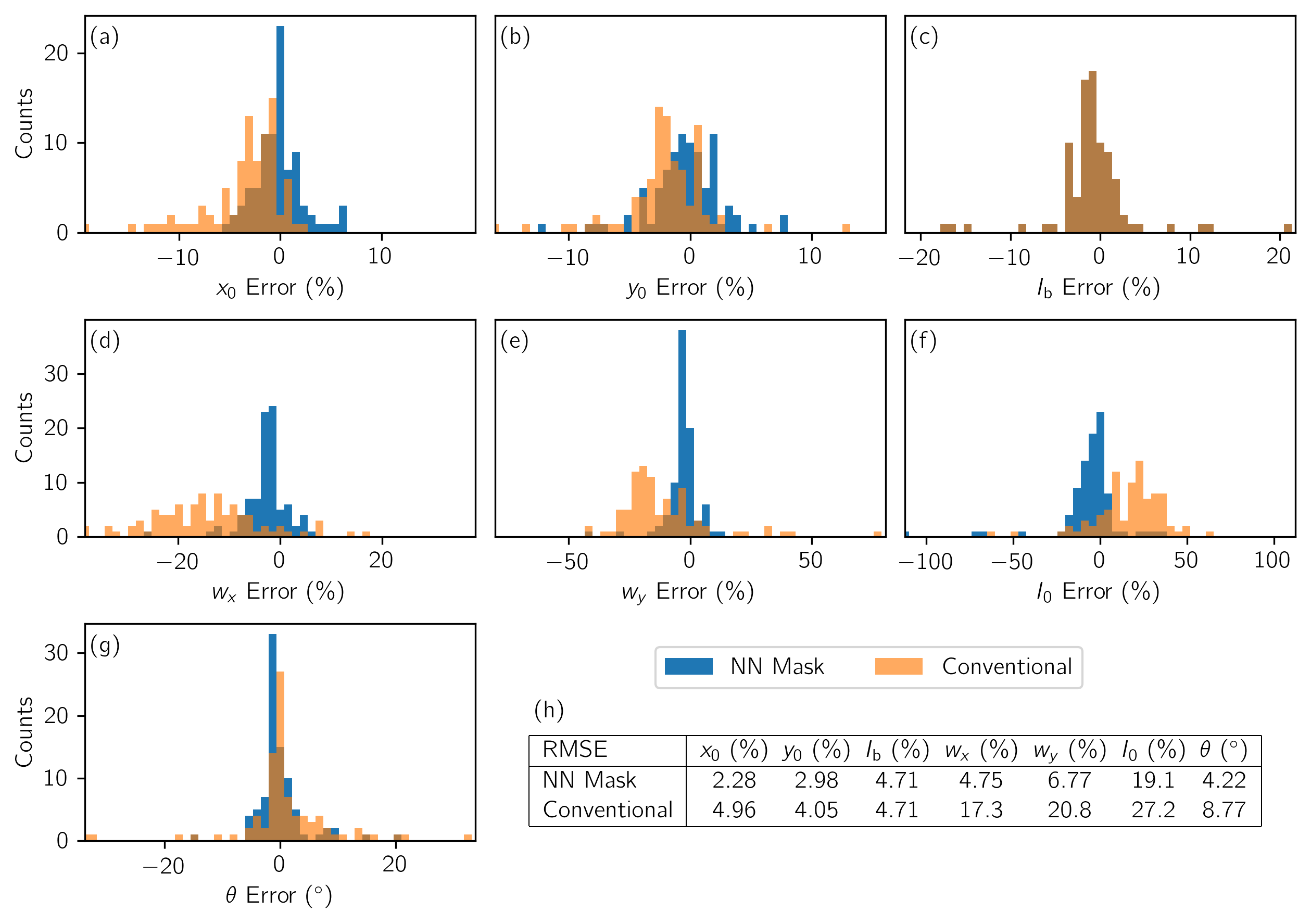}
\caption{Gaussian parameter errors. (a)--(g) Histograms of the extracted Gaussian parameter errors---calculated by comparison to the 2D fit values---for all the atom clouds in the validation dataset. The relative errors for the neural network segmentation mask method (blue) and the conventional method (orange) are both shown. Cloud center coordinates $\{x_0, y_0\}$ and cloud radii $\{w_x, w_y\}$ are normalized by the fitted cloud radii. The peak amplitude and the background intensity ($I_0$, $I_\text{b}$) are both normalized by the fitted peak amplitude. (h) The root-mean-square error (RMSE) of the Gaussian parameters.}
\label{fig:error}
\end{figure}

We can extract the parameters characterizing the cloud's 2D Gaussian distribution (see Eq.~\ref{eq:2dgauss}) directly from the NN's segmentation mask output (see Fig.~\ref{fig:flow}a). Taking the 1$^{\rm st}$ moments of the mask yields the center coordinates $\{x_0, y_0\}$ of the atom cloud, whereas the 1/$e^2$ radii $\{w_x, w_y\}$ and the angular orientation $\theta$ can be determined by calculating the 2$^{\rm nd}$ moments \cite{Hofer2017} of the mask. This method of extracting clouds' Gaussian parameters directly from the segmentation mask was applied to the validation dataset (see Fig.~\ref{fig:predict}c) and the results compared to the Gaussian parameters extracted via a 2D fit of the experimental image (see Fig.~\ref{fig:error}, blue bars). Apart from $\theta$, we normalized the differences to obtain a relative error by dividing the cloud center $\{x_0, y_0\}$ and the cloud radii $\{w_x, w_y\}$ by the fitted radii.

For comparison, we also extracted the Gaussian parameters from the clouds using a conventional non-fitting method (see Fig.~\ref{fig:flow}b); instead of getting $\{x_0, y_0, w_x, w_y, \theta \}$ directly from the NN's segmentation mask they are obtained by applying the  1$^{\rm st}$ and 2$^{\rm nd}$ moments to the experimental images within the ground truth ROIs. This method was applied to the images in the validation dataset with the extracted parameters then normalized and compared to those from the 2D fit, deemed the true parameters for the cloud, to calculate the relative error (see Fig.~\ref{fig:error}, orange bars).

For both methods, the background intensity $I_{\text{b}}$ is calculated by generating a histogram of the experimental image's pixel values (where there is no cloud) and then taking the peak position (see Fig.~\ref{fig:flow}c). The amplitude of the 2D Gaussian $I_0$ is calculated by summing the image's intensity inside the 1/$e^2$ contour to find the power ($P$) and then applying 
\begin{equation}
I_0 =\frac{2}{\left(1-e^{-2}\right)}\left(\frac{P}{\pi w_xw_y}-I_{\text{b}}\right)\text{.}
\label{eq:power}
\end{equation}
Since the peak amplitude is highly dependent on $\{x_0, y_0, w_x, w_y, \theta \}$, this results in significant differences for $I_0$ when calculated with the parameters extracted from the NN segmentation mask or those calculated using the conventional method (see Fig.~\ref{fig:error}f). The peak amplitude $I_0$ and the background intensity $I_{\text{b}}$ were both normalized when calculating the error by dividing by the fitted peak amplitude.

We qualitatively compare the two methods by calculating the root-mean-square error (across the validation image set) for each parameter (see Fig.~\ref{fig:error}h). Even though the conventional method is given the ground truth ROIs (which the NN method finds on its own), it is significantly outperformed by the NN segmentation mask method. The Gaussian parameters extracted from the NN's segmentation mask can either be used directly or as seed parameters for a conventional 2D fit---increasing the likelihood of fit convergence and reducing the fitting time---with a 30\% speed-up seen for our validation dataset compared to a fit without seed parameters.

\section{Conclusion}

An instance segmentation neural network (Mask R-CNN) was trained to identify ultracold atom clouds in magneto-optical traps and optical dipole traps. The neural network (NN) generates both a region-of-interest (ROI) and a segmentation mask for each cloud---corresponding to the cloud's 1/$e^2$ radii---with a mean average precision of 86.3\% and 85.8\% on the ROI and the segmentation mask branches, respectively. We show that the Gaussian parameters describing the atom clouds' distributions can be extracted directly from the segmentation masks with significantly higher accuracy than a conventional non-fitting method.

With an appropriate training dataset these techniques could be applied to ultracold atom clouds in other traps such as optical lattices \cite{bloch2005ultracold, viebahn2019matter} and box potentials \cite{Gaunt2013}; they are also directly applicable to laser beam profiling and other machine vision applications which require analysis of one or more 2D Gaussian distributions.

In the future a custom NN could be created by adding a branch after the ROI alignment stage which would output the cloud parameters directly. This would enable the characterization of non-Gaussian density profiles, useful, for example, in the detection, identification and parameterization of the bimodal clouds seen when a Bose--Einstein condensate \cite{anderson1995observation, davis1995bose} is present.

\section*{Acknowledgements and Funding}
We thank Elliot Bentine, Shu Ishida and Jirka Ku\v cera for helpful discussions and comments on the manuscript. This work was supported by EPSRC Grant No. EP/P009565/1, the John Fell Oxford University Press (OUP) Research Fund and the Royal Society.

\section*{Disclosures}

The authors declare no conflicts of interest.


%% file: hyperparameters.tex
\begin{tabular}{l @{\hspace{2\tabcolsep}} c @{\hspace{2\tabcolsep}} c @{\hspace{2\tabcolsep}} c @{\hspace{2\tabcolsep}}  c}
\hline \hline
Parameters & Lower Bound & Upper Bound & Log Scale & Best Value \\
\hline
Learning Rate &      0.0001 &       0.009 &       Yes &     0.0033 \\
Momentum      &         0.7 &        0.92 &        No &       0.86 \\
LRS Step Size &           3 &          15 &        No &         13 \\
LRS Decay     &       0.001 &           1 &       Yes &      0.049 \\
\hline \hline
\end{tabular}

%% file: output.bbl
\providecommand{\newblock}{}